\documentclass{appolb}

\usepackage{graphics}
\usepackage{graphicx}
\usepackage{epsfig}
\usepackage{color}
\usepackage{hyperref}
\usepackage{latexsym}
\usepackage{amsmath}
\usepackage{amsthm}
\usepackage{amsfonts}
\usepackage{amssymb}
\usepackage{dsfont}
\usepackage{bm}
\usepackage{graphics,psfrag}
\usepackage{graphicx,psfrag}
\usepackage[tight]{units}

\newcommand{\eq}[1]{(\ref{#1})}
\newcommand{\fig}[1]{Fig.~{\ref{#1}}}
\newcommand{\tab}[1]{Tab.~{\ref{#1}}}

\newcommand{\SU}[1]{\mathrm{SU}(#1)}

\renewcommand{\Re}{\mathfrak{Re}\,}
\newcommand{\tr}[1]{\,\mathrm{tr}\left[#1\right]}
\newcommand{\tmin}{t_{\mathrm{min}}}
\newcommand{\tmax}{t_{\mathrm{max}}}

\begin{document}
\title{Gauge fixing in lattice QCD with multi-GPUs%
\thanks{Presented at \emph{Excited QCD 2013}, Bjelasnica Mountain, Sarajevo.}%
}
\author{Mario~Schr\"ock\thanks{Speaker}
\address{Institut f\"ur Physik, FB Theoretische Physik, Universit\"at
Graz, A--8010 Graz, Austria}
\\\vspace{1.5em}
Hannes~Vogt
\address{Institut f\"ur Theoretische Physik, Auf der Morgenstelle 14, 72076 T\"ubingen, Germany}
}
\maketitle
\begin{abstract}
Here we present the \emph{cuLGT}\footnote{\url{www.cuLGT.com}} code for gauge fixing in lattice gauge field theories
with graphic processing units (GPUs).
Implementations for SU(3) Coulomb, Landau and maximally Abelian gauge fixing are
available and the overrelaxation, stochastic relaxation and simulated annealing 
algorithms are supported.
Performance results for single and multi-GPUs are given.
\end{abstract}
  
\section{Introduction}
Gauge fixing in lattice QCD is necessary in order to, e.g., compare lattice results to continuum
physics in a given renormalization scheme at a given scale.
The popular Landau gauge requires the four dimensional gradient of
the gauge field to vanish at each space-time point of the lattice.
The latter continuum condition translates to a large scale optimization problem 
in lattice gauge field theories.
Finding its maxima is very expensive in terms of computational costs and
a possible acceleration by high performance fine grained parallel architectures, like
graphic processing units (GPUs), is highly desirable.
Here we present a code written in CUDA which has been developed for the purpose of
lattice gauge fixing on GPUs.
The code makes strong use of template classes and algorithm abstraction to increase
its flexibility and to extend its applicability to related problems in lattice gauge
field theory.

Lattice QCD gauge fixing on GPUs was first presented in \cite{Schrock:2011hq}
and a detailed discussion of our code can be found in Ref.~\cite{Schrock:2012fj}.
The authors of \cite{Cardoso:2012pv} use the Fourier accelerated deepest descent
method for gauge fixing in lattice QCD.

In the following discussion we restrict ourselves to the example of Landau gauge fixing
and we refer to \cite{Schrock:2012fj} for the details
of the other gauges and algorithms which are supported by \emph{cuLGT}.

\section{Lattice Landau gauge}
The continuum Landau gauge condition,
\begin{equation}\label{eq:landaugauge}
	\partial_{\mu} A_\mu(x)=0,
\end{equation}
is fulfilled if and only if the lattice gauge functional
\begin{equation}\label{eq:landau_functional}
	F^g[U] = \frac{1}{N_cN_dV}\Re\sum_{\mu, x} \tr{U^g_\mu(x)},
\end{equation}
resides in a stationary point
with respect to gauge transformations $g(x)\in\mathrm{SU}(N_c)$. 
Here we denoted a gauge transformation of the link variables as
\begin{equation}
 U^g_\mu(x) \equiv g(x) U_\mu(x) g(x+\hat\mu)^\dagger.
\end{equation}
$N_c$ is the number of colors, $N_c=3$ for QCD,
$N_d$ is the number of space-time dimensions, (here $N_d=4$) and
$V$ is the total number of lattice points.

A measure $\theta$ of the Landau gauge precision
is the average $L_2$-norm of the gauge fixing violation $\Delta(x)$, i.e.,
the discrete derivative of the continuum gauge fields
\begin{equation}
	\Delta(x)\equiv \sum_\mu\left(A_\mu(x)-A_\mu(x-\hat \mu) \right) =0,
\end{equation}
\begin{equation}
\theta\equiv \frac{1}{N_cV}\sum_{x}\tr{\Delta(x)
\Delta(x)^\dagger}.
\end{equation}

\section{The relaxation algorithms}
The idea of the relaxation algorithms is to sweep over the lattice site by site
while optimizing the gauge functional locally. All 
sites of one of the two \emph{parity subsets} (checker board decomposition) 
can be optimized at the same time
because the newly generated local optimum depends on the nearest neighbors only.

Instead of taking the complete global gauge functional into account,
\begin{equation}\label{eq:landau_loc}
	F^g[U] = \frac{1}{2N_cN_dV} \Re\sum_{x} f^g(x),
\end{equation}
the relaxation algorithm aims at optimizing the value of $F^g[U]$ locally, i.e., we search 
the maximum of 
\begin{equation}
	f^g(x) = \Re\tr{g(x)K(x)}
\end{equation}
for all $x$. Here we defined
\begin{equation}\label{Kx}
	K(x):= \sum_\mu\Big( U_\mu(x) g(x+\hat\mu)^\dagger 
		+ U_\mu(x-\hat\mu)^\dagger g(x-\hat\mu)^\dagger\Big)\,.
\end{equation}
For $\mathrm{SU}(2)$, the maximum thereof is given by
\begin{equation}\label{localsolution}  
g(x) = K(x)^\dagger/\sqrt{\det{K(x)^\dagger}}
\end{equation}
and for $\mathrm{SU}(3)$  one iteratively operates in the 
three $\mathrm{SU}(2)$ subgroups
\cite{CabibboMarinari1982} and thereby optimizes the local $\SU{3}$ gauge functional.

\subsection{Overrelaxation}
Replacing the local gauge transformation $g(x)$ by $g^\omega(x),\;\omega\in[1,2)$ 
reduces the \emph{critical slowing down} of the relaxation algorithm on large
lattices \cite{Mandula:1990vs}.
In practice the exponentiation of the gauge transformation is done to first order.

\subsection{Stochastic relaxation}
The stochastic relaxation algorithm replaces the local gauge update $g(x)$
by  $g^2(x)$ with probability $p$ and
can speed up the convergence on large lattices.

\section{Single-GPU implementation}
\begin{figure}
	\center
	\includegraphics[width=1.0\textwidth]{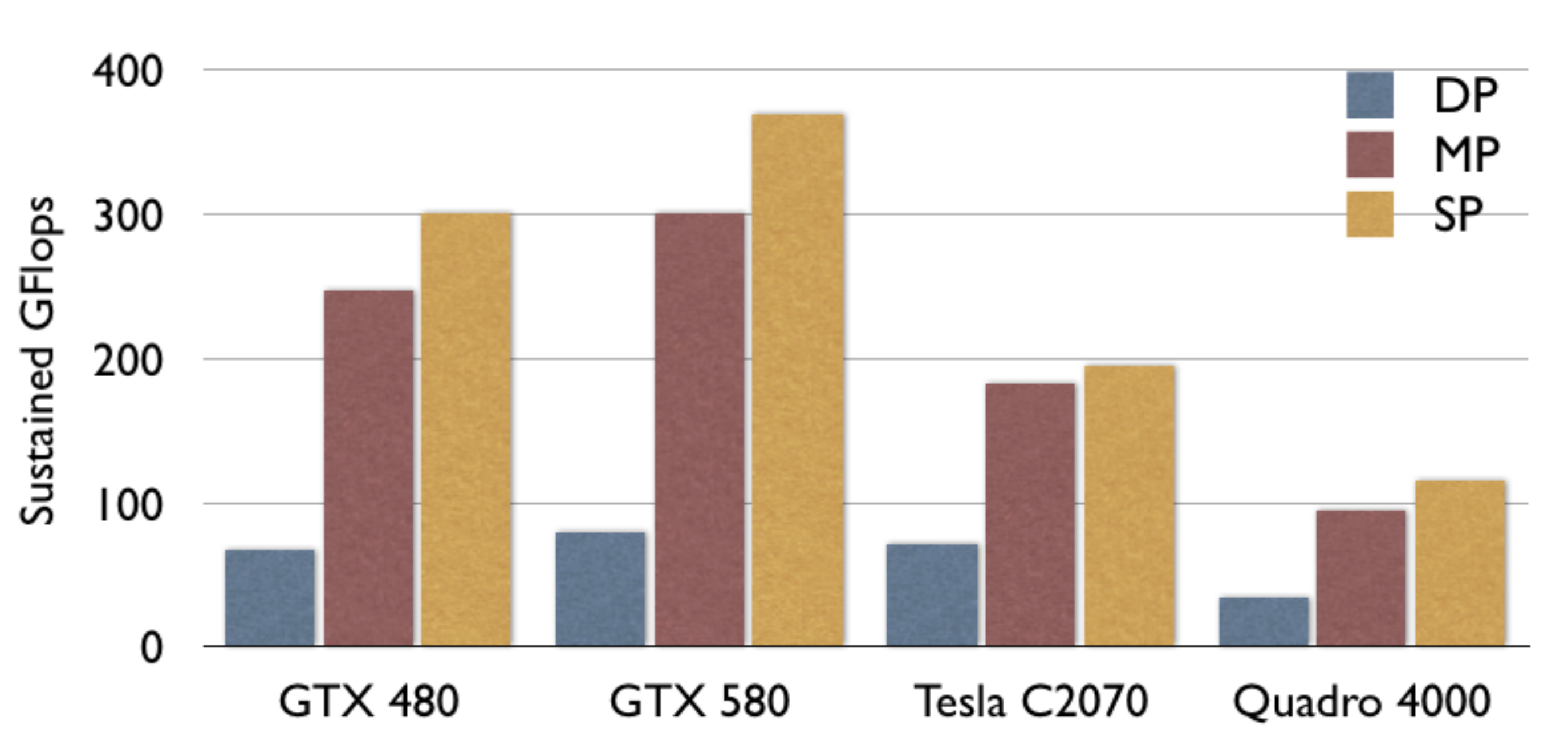}
  \caption{Performance of the overrelaxation kernel on different NVIDIA devices in single (SP), mixed (MP) and double precision (DP) on a $32^4$ lattice.
}\label{fig:devicehisto}
\end{figure}

We assign eight CUDA threads to each lattice site of a given parity in order to
calculate and apply the local gauge update \eq{localsolution}.
The two parity sublattices are treated consecutively and the relaxation algorithm
is iterated until the requested gauge precision $\theta$ has been reached.
A variable data storage pattern for the gauge fields is adopted in order to meet
the memory coalescing constraints of the hardware.
In \fig{fig:devicehisto} we compare the performance of the code on different
NVIDIA devies.

\section{Multi-GPU implementation}
\begin{figure}
	\center
	\includegraphics[width=1.0\textwidth]{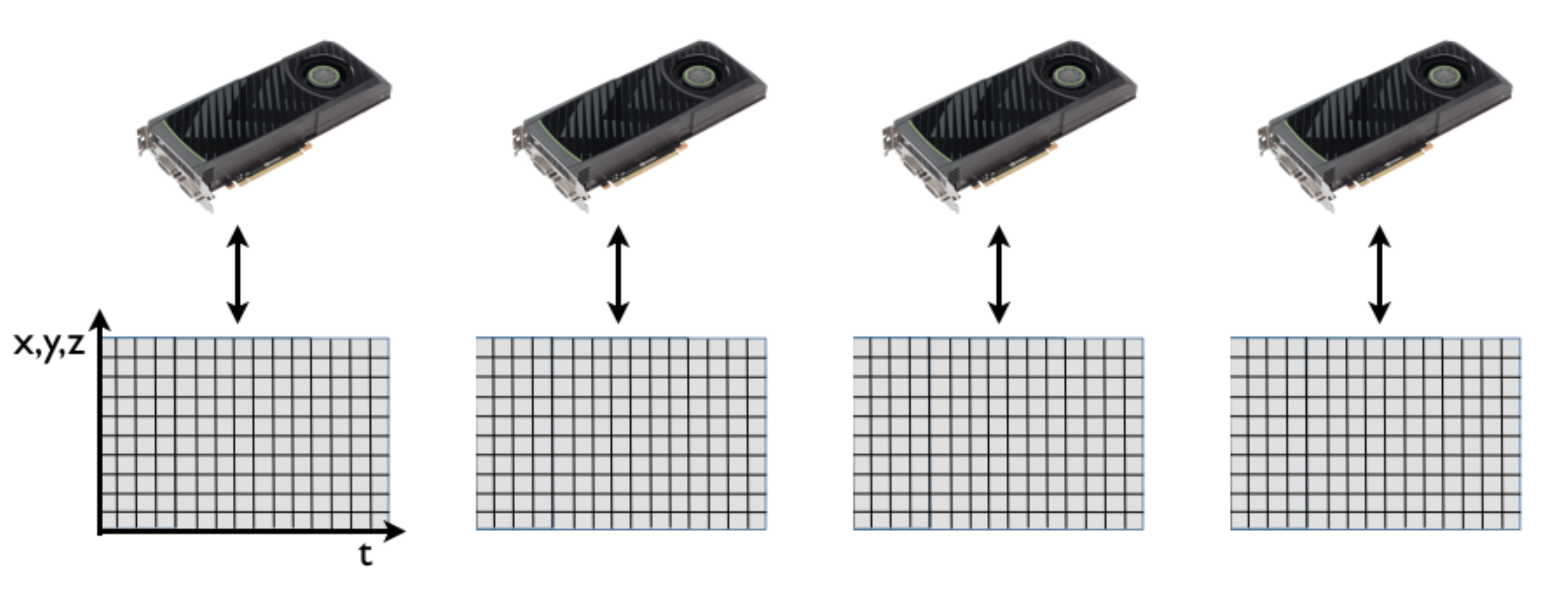}
  \caption{The four dimensional lattice is split along the temporal axis and
  distributed to the devices. The bottle neck is the communication at the boundaries
  via the PCI-bus.
}\label{fig:domaindecomp}
\end{figure}

For the multi-GPU implementation we decided for a decomposition of the lattice along the
temporal axis, see \fig{fig:domaindecomp}.
In each step of the iteration, the gauge links of the neighbor device in the temporal direction
have to be exchanged via MPI in order to calculate the gauge update \eq{localsolution}.
After the gauge update has been calculated it has to be applied to all connected gauge links, 
therefore it has to be copied to the neighbor device.
In detail, 
the following set of instructions has to be carried out on each device
in order to transfer the links $U_0(\tmax)$ of device $i$ to device $i+1$:
\begin{enumerate}
\item \emph{cudaMemcpyDeviceToHost} of $U_0(\tmax)$ (inactive parity)
\item \emph{MPI\_Send} of $U_0(\tmax)$ to device $i+1$ and
	\emph{MPI\_Recv} of $U_0(\tmin-1)$ from device $i-1$
\item \emph{cudaMemcpyHostToDevice} of $U_0(\tmin-1)$
\item update $U_\mu(\tmin)$ (active parity) which affects $U_0(\tmin-1)$ (inactive)
\item \emph{cudaMemcpyDeviceToHost} of $U_0(\tmin-1)$ (inactive parity)
\item \emph{MPI\_Send} of $U_0(\tmin-1)$ to device $i$ and
	\emph{MPI\_Recv} of $U_0(\tmax)$ from device $i+1$
\item \emph{cudaMemcpyHostToDevice} of $U_0(\tmax)$ 
\end{enumerate}

In order to hide to slow data exchange over the low-bandwidth PCI-bus,
we perform asynchronous memory transfers: we overlap the data exchange on the
boundaries with calculations in the inner part of the domain.
\begin{table}
	\center
	\begin{tabular}{c|c|c|c|c|c}
	$N_s^3$ & D2H [$\mu s$] & H2D [$\mu s$] & kernel [$\mu s$] &
	 D2H/kernel & H2D/kernel \\\hline
		16  &  0.0398  & 0.0368 & 0.0209 & 1.90 & 1.76 \\
		32  &  0.2543  & 0.2276 & 0.1443 & 1.76 & 1.58  \\
		64  &  1.2510  & 1.1830 & 1.0489 & 1.19 & 1.13  \\
		128 &  8.9597  & 8.7169 & 8.3041 & 1.08 & 1.05
	\end{tabular}
	\caption{Time in microseconds needed to copy the data at the boundaries
	from device to host (D2H) and host to device (H2D)
	compared to the time needed to update one time-slice with the 
	overrelaxation kernel. The two most right columns
	give the ratios.
	}
	\label{tab:data_exchange}
\end{table}
In \tab{tab:data_exchange} we compare the time needed to update one time-slice in the
inner part of the domain with the time needed to copy the data at the boundaries to
the host memory and from the host memory to the neighboring device.

\begin{figure}
	\center
	\includegraphics[width=0.49\textwidth]{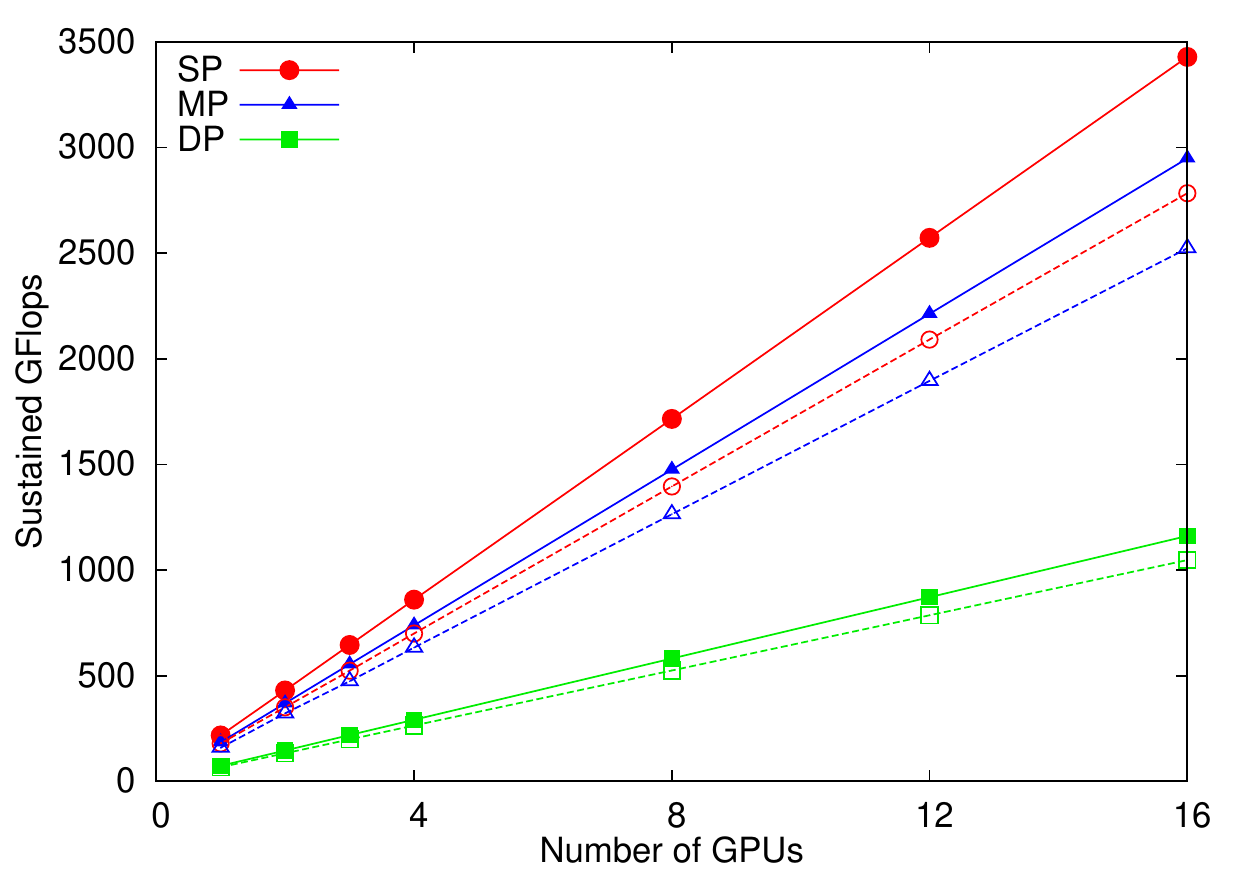}
	\includegraphics[width=0.49\textwidth]{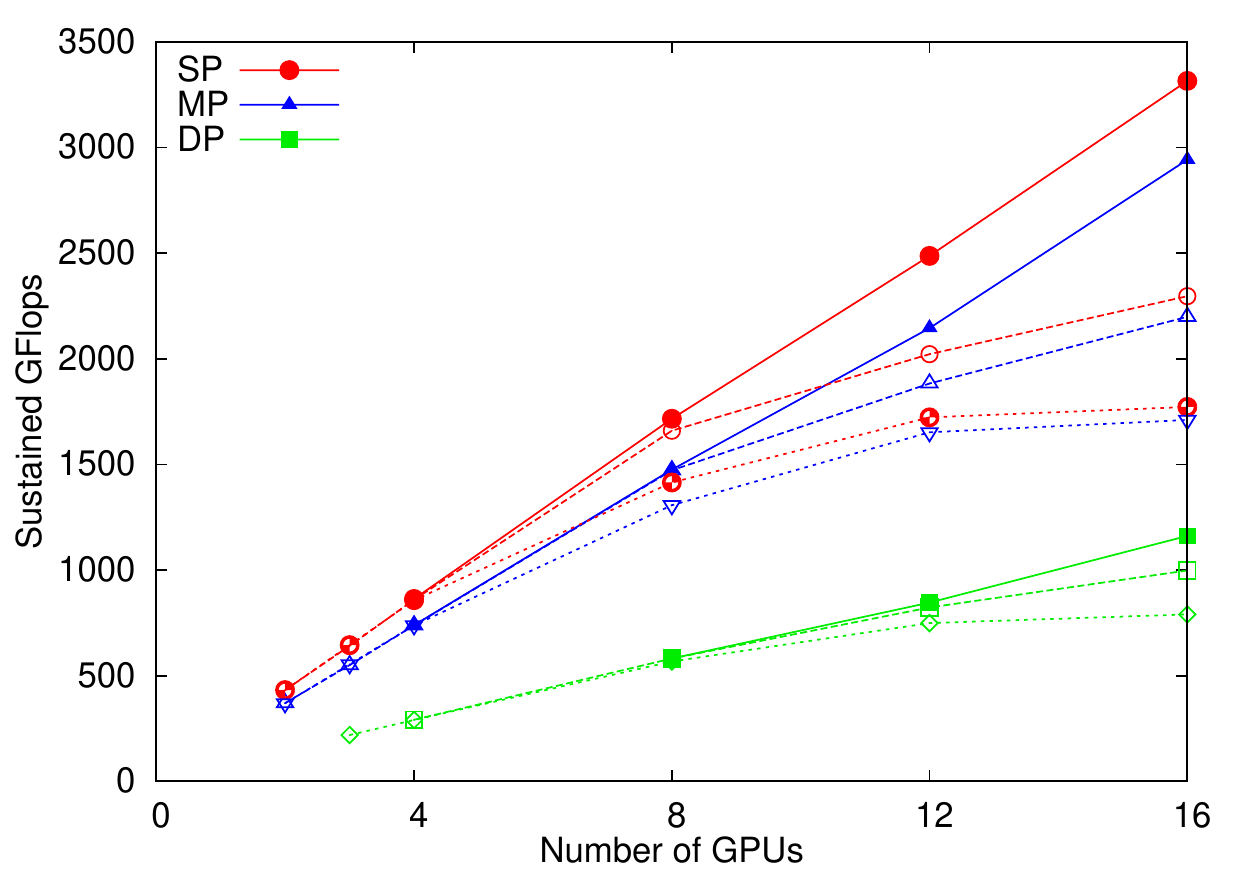}
  \caption{Left: weak scaling on the NVIDIA Tesla C2070.
		The full symbols correspond to a lattice size of $64^3\times32$ per GPU and the
		open symbols to $48^4$ per GPU.
		Right: strong scaling on the Tesla C2070. 
		The spatial lattice volume is kept fixed at $64^3$ and the total temporal extent
		varies for the three lines (per precision) from the top downwards
		$N_t=256, 128, 96$.
}\label{fig:scaling}
\end{figure}

\fig{fig:scaling} confirms the predictions of \tab{tab:data_exchange} that linear
weak scaling is achieved with asynchronous memory transfers.

\section{Summary}
The local relaxation algorithms for lattice gauge fixing are well suited to be 
accelerated with highly parallel architectures like GPUs.
With the aim of retaining maximum performance in a multi-GPU implementation
it is crucial to overlap the data exchange between the devices by calculations
in the inner part of the domain.
This allows for linear weak scaling and a maximum performance of $\sim 3.5$ Teraflops
adopting 16 Tesla~C2070 GPUs.

\section*{Acknowledgments}
Support by the Research
Executive  Agency (REA) of the European Union under Grant Agreement 
PITN-GA-2009-238353 (ITN STRONGnet) 
and by the Austrian Science Fund (FWF) through grant DK W1203-N16
is gratefully acknowledged.



\end{document}